# Principle of least effort vs. maximum efficiency: deriving Zipf-Pareto's laws

Qiuping A. Wang

Laboratoire Systèmes Complexes et Information Quantique

ESIEA, 9 Rue Vésale, 75005 Paris

IMMM, CNRS UMR 6283, Le Mans Université, Ave. O. Messiaen,

72085 Le Mans Cedex, France

## Abstract

This paper provides a derivation of Zipf-Pareto laws directly from the principle of least effort. A probabilistic functional of efficiency is introduced as the consequence of an extension of the nonadditivity of the efficiency of thermodynamic engine to a large number of living agents assimilated to engines, all randomly distributed over their output. Application of the maximum calculus to this efficiency yields the Zipf's and Pareto's laws.

Keywords: Least effort, Maximum efficiency, Zipf's law, Pareto law, calculus of variation

PACS codes: 89.75.−k Complex systems; 89.75.Da Systems obeying scaling laws; 02.70.Rr General statistical methods; 02.30.Xx Calculus of variation

## 1) Introduction

Zipf's law is an empirical law describing the discrete distribution of some measured values $x$ as a function of their rank $r$. If all the measured values are binned into $W$ rank r in a decreasing order $(x_1 > x_2 > x_3 .... > x_r ... > x_W)$, Zipf's law is given by [1][2]

$$x_r = \frac{x_1}{r^\alpha} \qquad (1)$$

where $\alpha$ is a parameter to be determined experimentally. This regularity was first discovered by Estoup [3] and 30 years later by Zipf [1][2] who popularized it through more systematic and wider investigation of event frequency distribution. For language systems, $x_r$ represents the frequency of word in given text, $\alpha$ being often close to one [2], but also possibly very different (larger or smaller) from unity in different language systems [4]. Other examples of rank distributions include, for instance, city size, where $x_r$ is population of a city with $\alpha$ between 0.5 and unity [6], and individual income distribution, where $x_r$ is individual income with $\alpha$ between 0.5 and 1.5 ([2], [6] and references therein). In many cases, Zipfian law as a straight-line with a unique slope in log-log plot is just an approximate representation of the decreasing tendency in the rank distribution [13][24][25], better representations include, among many others, the Mandelbrot formula with one more adjustable parameter [18] and the generic models described in [13][24][25][26]. In my opinion, what is essential in all those observed Zipf or near-Zipf laws is the decreasing tendency of rank distributions, more or less close to a straight line in log-log plot, which must have some underlying universal mechanisms, being generic and common to all (living) systems considered.

Zipf gave a first interpretation of this decreasing rank distribution law by postulating that all human beings minimize effort in their activities to get some fulfillment [2]. This rule was first formulated in 1894 by Ferrero in his paper discussing the mental inertia of human being [5]. But Zipf was the first to explore its possible application to quantitative study. He wrote : *"The **power laws** in linguistics and in other human systems reflect an economical rule: everything carried out by human being and other biological entities must be done with least effort (at least statistically)"*. This rule was obviously an intuition from the observation of the behaviors of human being himself and probably of other animals, always trying to get more done by doing less. A summary is given in [6] about several experimental works checking the rule of least effort.

The idea of least effort for human and animal systems is very appealing and even fascinating, especially in a perspective of using it for quantitative and analytic methods in the same way as many variational principles in physics (stationary action, least time, maximum entropy etc.). However, no such implementation with convincing mathematical calculus has been realized to date to our knowledge. The reader can find a summary of some of the efforts in [6], of which it is worth mentioning the work of Mandelbrot [18] and Canchu *et al* [21] who defined functionals of effort by considering information in language communication in order to apply the idea of least effort and to derive Zipf law for languages. Their work is an important step towards the implementation of the principle of least effort. I would only give two remarks here about the limits of their approach. The first is that the expressions of effort (cost) of [18], proportional to log of rank, is a little ad hoc, since it is obviously suggested in order for the variational calculus of maximum Shannon information to yield Zipfian power law, while the conventional outcome of this calculus is exponential distribution law. Another questionable point of [18] has been remarked by Manin [22] about the mutual dependence of the two parameters of the model which were thought to be independent. Concerning the proposition of Canchu et al [21], the functional of effort is proportional to the information communicated between speaker and hearer, while in a communication process, the information transferred should be the output of effort instead of effort itself. The effort in this case should be process dependent. If you are buying information, your effort is money. If you are searching for information in documents, your effort is related to time spent or to the volume of documents read. If you are trying to find information on internet, your effort can be estimated by the number of clicks made. If you are expressing information by speaking or writing, your effort is related to the number of different words employed or to the total number of words, and so on. Obviously, the quantity of information you obtain is not necessarily proportional to your effort (money, time, clicks, number of words etc.). My second remark is that the approaches of [18] and [21] are limited to language and cannot be used to implement least effort for different living systems.

So the relationship between this beautiful principle and the Zipf's law remains a sort of speculation to date without solid and convincing mathematical proof. In the past several decades, much attempt was made to interpret or derive this power law with different mechanisms and models. There are almost as many models proposed as the systems in which Zipf's law and near-Zipf's laws have been observed. The reader can easily find relative information. The most recent models include, to our knowledge, about the origin of the Zipf's law in language by considering the interaction between syntax and semantics [7].

Another law we will focus on in this work is the Pareto distribution, a power law originally describing the wealth distribution of a population in a given society [8]:

$$P(X > x) = \left(\frac{x_m}{x}\right)^{\beta} \qquad (2)$$

where $X$ is a random variable representing the income, $P(X > x)$ the probability of finding a person with income larger than a value $x$, $x_m$ the smallest income and $\beta$ a constant characterizing the distribution. This distribution law is the origin of the famous 20-80 rule of Pareto. It is believed that there is an intrinsic link between Zipf's and Pareto's laws. If one is observed in a system, another must exist simultaneously [6][19]. There is also an opposite view [20] which I will discuss later on in this paper

The aim of this work is to derive these laws representing the decreasing tendency in the distributions, in a generic way from the idea of least effort for living systems or systems driven by living agents. For this purpose, a universal functional of effort is necessary for the minimum calculus of variation. However, an effort is a cost whose nature differs in different domains and processes. It can be an expenditure of energy, time, information, an amount of money and so on. It is difficult to define and quantify an effort in a general manner. In this work, I focus on another relative quantity instead: the efficiency. The notion of energy conversion efficiency is used throughout this work. This efficiency is defined by the ratio of the useful energy delivered by a system to the energy supplied to it over the same period of operation[1]. If a living agent, making effort for some achievement (output), is likened to the system delivering useful energy, its effort can be assimilated to the input (supplied) energy. We can then maximize the efficiency instead of minimizing effort. The maximum efficiency (MAXEFF) is obviously more advantageous and general than minimum effort, since, as defined above, efficiency involves both effort and output, its maximization implies not only minimum effort, but also maximum output which would be absent if the variational calculus of least effort was made with only a functional of effort. To my opinion, the true spirit of the principle of least effort can only be implemented by MAXEFF calculus.

---

[1] https://www.energy.ca.gov/resources/energy-glossary

The idea of MAXEFF in science and engineering is not new. A good example is the derivation of the Betz limit of the efficiency of wind turbine from fluid mechanics[2] [9]. The essential of the application of MAXEFF is to use an expression of efficiency as a functional. I adopt this method and introduce a functional of efficiency by considering the nonadditive property of the thermodynamic efficiency and the fact that we are tackling a large number of engines (living agents), all distributed over some output. Then the application of the maximum calculus to this functional of probability distribution yields the Zipf's and Pareto laws.

**2) The Nonadditivity of efficiency**

The definition of efficiency of a thermal engine in thermodynamics differs from one type of engines to another. For example, suppose an work engine absorbs an energy $Q_1$, produces a useful work $W$, and rejects an energy $Q_2$, in the ideal case without energy loss where all heat cost $Q_1 - Q_2$ is converted into work $W = Q_1 - Q_2$, the efficiency of this engine is defined by

$$\eta = \frac{W}{Q_1} = 1 - \frac{Q_2}{Q_1} \tag{3}$$

For a heat pump engine (conditioner for heating for example) which absorbs a heat $Q_1$, consumes a work $W$, and produces a heat $Q_2$ for heating. We have $W = Q_2 - Q_1$ if all work is converted into heat. Its efficiency is defined by $\eta = \frac{Q_2}{W} = \frac{1}{1 - \frac{Q_1}{Q_2}}$ or

$$\frac{1}{\eta} = \frac{W}{Q_2} = 1 - \frac{Q_1}{Q_2} \tag{4}$$

For a refrigerator (conditioner for cooling for example) which absorbs a heat $Q_1$, consumes a work $W$, and rejects a heat $Q_2$ for cooling, we have $W = Q_2 - Q_1$ if all work is converted into heat. Its efficiency is defined by $\eta = \frac{Q_1}{W} = \frac{1}{\frac{Q_2}{Q_1} - 1}$ or

$$\frac{1}{\eta} = \frac{W}{Q_1} = \frac{Q_2}{Q_1} - 1 \tag{5}$$

The nonadditivity relationships of $\eta$ and of $\frac{1}{\eta}$ are similar (see calculation in the Appendix I). In what follows, I only give a summary of the nonadditivity of $\eta$ for two working engines. Suppose two engines are connected in such a way that the first engine absorbs an energy $Q_1$, does a work $W_1$, and rejects an energy $Q_2$, and the second engine absorbs an energy $Q_2$, does a work $W_2$, and rejects an energy $Q_3$, one has $\eta_1 = \frac{W_1}{Q_1} = 1 - \frac{Q_2}{Q_1}$, $\eta_2 = \frac{W_2}{Q_2} = 1 - \frac{Q_3}{Q_2}$. The overall efficiency $\eta$ of the ensemble of two engines is defined by $\eta = \frac{W_1+W_2}{Q_1} = 1 - \frac{Q_3}{Q_1}$. It is straightforward to calculate

$$\eta = \eta_1 + \eta_2 - \eta_1 \eta_2. \tag{6}$$

---

[2] https://www.youtube.com/watch?v=U_195_Uq3dU, https://en.wikipedia.org/wiki/Betz%27s_law

If the engines cannot transform all the heat cost ($Q_1 - Q_2 = W_1$ for the first engine for example) due to energy loss (friction, vibration, heat conduction, heat radiation and so on), we can introduce a loss coefficient $a$ in such a way to write (see Appendix I)

$$\eta = \eta_1 + \eta_2 + a \eta_1 \eta_2. \tag{7}$$

Formally, $a = -1$ corresponds to the case where the energy cost is totally converted into work. If $a < -1$, this is the case where the engines cannot transform all the heat cost into useful work, leading to a reduction of efficiency with respect to the case of $a = -1$. On the contrary, if $a > -1$, there is an enhancement of efficiency as if the collaboration created energy with respect to the $a = -1$ case. This is possible only for processes in which the input and output are not energy or energy proportional quantities. This point will be discussed in the following section.

It is noteworthy that Eq.(7) takes place only for two engines with the output of one being the input of another. Different coupling between the engines gives rise to different additivity. For example, if the second engine does not consume all the heat $Q_2$ rejected by the first engine, another parameter is necessary for describing this partial collaboration (see Appendix I-A2). For two engines in contact with the same heat baths but having no exchange of energy, the total efficiency depends on the efficiency of each engine in a much more complicated way. In this work, we address living systems made up of coupled living agents having exchange of energy, information or other resources, making it possible for the output of ones to be the input of others. This property has been widely observed in most biological and social systems, at least those showing Zipf-Pareto laws [6]. The choice of the nonadditivity of Eq.(7) in this work is motivated by this consideration.

### 3) Modeling an ensemble of thermal engines

Suppose an ensemble of a large number $N$ of engines (or living agents). They are independent in the sense that there is no direct interaction or correlation between them, but their functioning s cooperative in the sense that the output of one engine can be used as the input of others and vise versa. This exchange of something (energy, information, entropy, money, even time) does not necessarily the existence of interactive forces between the agents. We do not consider the case where engines have no communication or collaboration between them. Each engine has a certain efficiency $\eta_n$ with $n = 1, 2 \ldots N$. The total efficiency $\eta$ of the ensemble should be a function of all $\eta_n$: $\eta = f(\eta_1, \eta_2 \ldots \eta_N)$.

Efficiency is in general nonadditive; hence, the total one $\eta = f(\eta_1, \eta_2 \ldots \eta_N)$ cannot be a simple sum of the individual efficiencies. Based on the above analysis of the nonadditivity relationships for different type of engines (doing work, heating, cooling) and different collaboration type, we can model the whole system by using a simple nonadditivity given by Eq.(7) which also reads $(1 + a\eta) = (1 + a\eta_1)(1 + a\eta_2)$. This equation can be written as, for the whole system of $N$ engines:

$$(1 + a\eta) = \prod_{n=1}^{N}(1 + a\eta_n) \qquad (8)$$

where $a$ is a parameter that characterizes collaboration between the engines as well as the energy loss during the processes from input to output. For many systems, $a$ can be free from energy connection for non-thermodynamic processes for which the input and the output quantities are not energy connected and there is no necessarily energy conservation condition. For example, for living agents trying to be connected to some objects (sites, friends, cities, richness etc.), the output can be frequency of connection, the population or the agents' richness. There is no energy conservation between these quantities and the inputs which can be energy cost, expenditure of time or money, used materials, number of actions and so forth. Another example is the economic process of investment. This process is similar to the process of a heat engine. The invested amount of money can be assimilated to the input heat $Q_1$, the consumed input work $W$ is the effort to increase the profit, and the total turnover can be considered as the heat production $Q_2$. In thermodynamics there would be a conservation condition: $Q_2 - Q_1 = W$. But this relationship does not exist for the economic process of investment because there is no quantitative measure of the effort $W$ and of its conversion to $Q_2$.

Eq.(8) is obviously the simplest model for the efficiency as a nonadditive quantity. The mathematical advantage of this model will be shown later. A little bit more complicated model can be

$$(1 + a\eta) = \prod_{n=1}^{N}(1 + a_n \eta_n) \qquad (9)$$

where $a$ is the parameter characterizing the whole ensemble and $a_n$ is the parameter of the n[th] subsystem or agent. This composite model can be used when it is necessary to consider the composite effect of subsystems. In what follows, I focus on the ensemble as a whole, the one parameter model Eq.(7) or (8) will be used.

## 4) Efficiency as a functional of probability

Suppose that all agents in the ensemble are making effort to achieve as much as possible a measurable quantity represented by a variable $X$ having $w$ discrete values $x_i$ with $i = 1,2,...w$. More they get that quantity, larger is $x_i$. This quantity can be income, wealth, city population, firm size, frequency of events, information quantity and so forth. At equilibrium (or stationary) state of the whole systems, all agents are distributed over the range of $X$ with $n_i$ agents at the value $x_i$. We have $\sum_{n=1}^{w} n_i = N$. The probability $p_i$ of finding an agent at the value $x_i$ is $p_i = \frac{n_i}{N}$. The normalization condition is $\sum_{i=1}^{w} p_i = 1$.

Due to the statistical nature of the model with a large number of agents distributed over all the values of $X$, it is reasonable to suppose that the total efficiency $\eta_i$ of the agents that have the value $x_i$ depends on the number $n_i$ with $\eta_i = f(n_i)$ or on the probability distribution $\eta_i = f(p_i)$. The average efficiency $\eta$ of the whole system reads $\eta = \sum_{i=1}^{w} p_i \, \eta_i$.

Now let us separate the whole ensemble of agents into two *independent* subsystems $A$ and $B$, with efficiency $\eta_k(A)$ and $\eta_j(B)$, respectively. The probability distribution of the agents in $A$ is $p_k(A)$ and that in $B$ is $p_j(B)$. The probability distribution of the whole ensemble can be written as

$$p_i = p_k(A) p_j(B) \text{ with } i = k, j \qquad (10)$$

We choose Eq.(7) as the efficiency nonadditivity. This implies a total efficiency given by

$$\eta_i = \eta_k(A) + \eta_j(B) + a\eta_k(A)\eta_j(B) \qquad (11)$$

or $(1 + a\eta_i) = [1 + a\eta_k(A)][1 + a\eta_j(B)]$. It can be proved that Eq.(10) and Eq.(11) together lead uniquely to $(1 + a\eta_i) = p_i^b$ or

$$\eta_i = \frac{p_i^b - 1}{a}. \qquad (12)$$

Obviously $\eta_k$ and $\eta_j$ have the same functional of $p_k$ and $p_j$, respectively. The proof of the unicity of Eq.(12) is shown in Appendix II.

The parameter $b$ is related to $a$ by the following considerations. First, due to the fact that the efficiency $\eta_i$ is positive and that $p_i$ is smaller than unity, $b$ should have opposite sign of $a$. Secondly, from Eq.(11), if $a$ goes to the zero limit $a \to 0$, the efficiency tends to additive limit $\eta_i \to \eta_k(A) + \eta_j(B)$. Taking into account Eq.(10), we expect an asymptotic behavior $\eta_i \to \frac{b\ln p_i}{a} \to -\ln p_i$ and $b \to -a$ for $a \to 0$. The simplest relationship is $b = -a$. Other relations $b = f(a)$ may be possible with any odd function $f(a)$ obeying $\lim_{a \to 0} f(a) \to -a$. However, as is shown in the coming sections, only $b$ is associated with the indices of power law distributions, $a$ being absorbed in the normalization constant. Therefore, different relationships $b = f(a)$ do not affect the power law derivation from the efficiency. The behavior of the efficiency is not affected very much either, especially when $a$ or $b$ are limited in a small interval around zero (see below). With no loss of generality, I choose the simplest relation $b = -a$ leading to

$$\eta_i = \frac{p_i^{-a} - 1}{a} \tag{13}$$

which tends to $\eta_i = -\ln p_i$ as $a$ tends to zero, allowing the additive efficiency relationship $\eta_i = -\ln p_k p_j = -\ln p_k - \ln p_j = \eta_k(A) + \eta_j(B)$. Finally, the average efficiency of the whole ensemble of $N$ agents reads

$$\eta = \sum_{i=1}^{w} p_i \eta_i = \frac{\sum p_i^{1-a} - 1}{a} \tag{14}$$

in which the normalization $\sum p_i = 1$ is considered. From now on, if not specified, the summation is over all the $w$ possible values (states) of $X$ for the whole ensemble. It is noteworthy that the above discussion addresses the discrete case. The continuous case will introduce other conditions on the positivity of $\eta$ [10].

In order to see the variation of $\eta$, let us imagine a two states system with $i = 1,2$. $\eta = \frac{\sum_{i=1}^{2} p_i^{1-a} - 1}{a} = \frac{p_1^{1-a} + p_2^{1-a} - 1}{a}$. Let $p_1 = p$, $p_2 = 1 - p$ following the normalization. We obtain $\eta = \frac{p^{1-a} + (1-p)^{1-a} - 1}{a}$ plotted in **Figure 1**. $\eta(p)$ increases with increasing $a$ in the interval $\eta \in \,]0, \infty[$ for $a \in \,]-\infty, \infty[$, and is concave for $a<1$, constant at $a=1$ and convex for $a>1$ with its extremum at $p = \frac{1}{2}$. This is to be proved analytically in the following section.

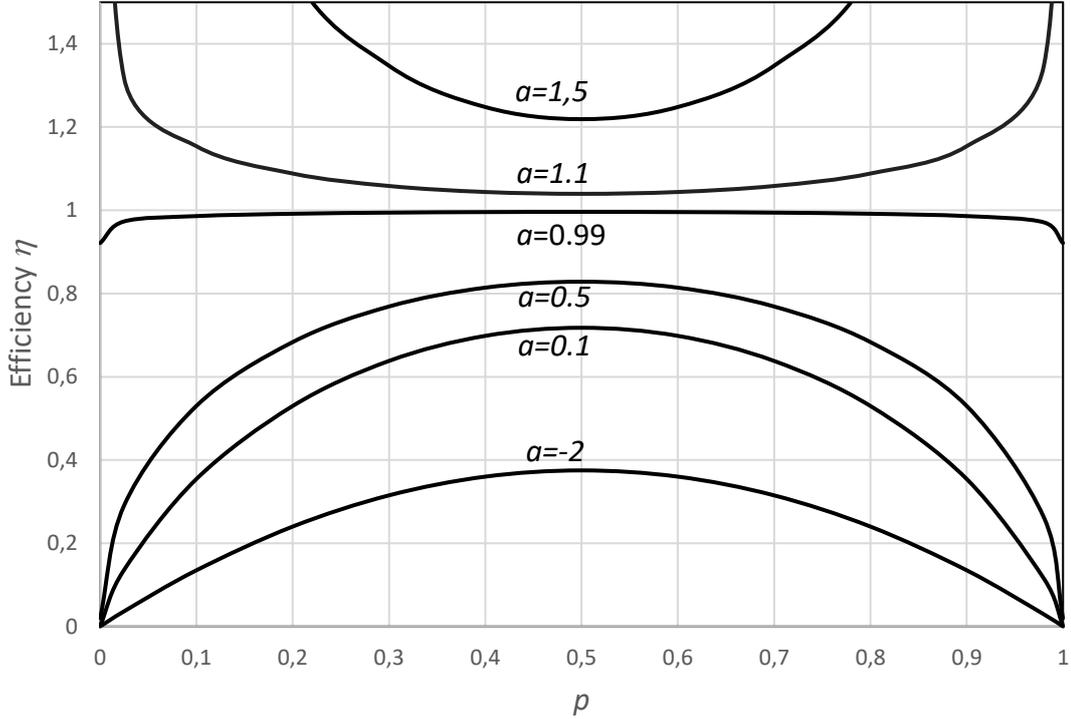

**Figure 1**: The variation of the efficiency as a function of the probability distribution over two states $p_1 = p, p_2 = 1 - p$ for different values of $a>0$. $\eta$ is concave for $a<1$, constant at $a=1$ and convex for $a>1$ with its extremum at $p = \frac{1}{2}$. It is noteworthy that $\eta$ is an monotonically increasing function of $a$ in the interval $0 < \eta < \infty$ for $-\infty < a < \infty$.

## 5) Maximization of efficiency

As mentioned in the introduction, following the idea of least effort, I propose to maximize the efficiency, meaning that a calculus of variation applied to the functional of the average efficiency in Eq.(14) with respect to the probability $p_i$ ($i = 1, 2 \dots w$). However, it is easy to verify that the maximization of $\eta$ alone cannot lead to correct probability distribution. Indeed, $\delta\eta = 0$ means $\frac{\partial \eta}{\partial p_i} = 0$ for all $i$, leading to $p_i^{-a} = 0$ meaning that $p_i = \infty$ for $a > 0$ and $p_i = 0$ for $a < 0$. The introduction of normalization as a constraint of the variation, i.e., $\delta(\eta + \sum p_i) = 0$, will lead to uniform distribution $p_i = 1/w$ for all $i$, which can be seen in **Figure 1**. This is of course not what we are looking for.

The maximization of efficiency is not an isolated property. The efficiency is at its maximum when fulfillment is the best due to the effort of the living agents. Hence, there must a connection between the efficiency and the fulfillment or the output. It is quite reasonable to associate the maximum of efficiency to the maximum output. The output is represented by the variable $X$. As a consequence, its average, $\bar{X} = \sum p_i x_i$, should be maximized at the same time as the efficiency. As mentioned above, $X$ can be the income, the frequency of events, the population of cities, the size of companies representing their wealth and so forth. In the case of income for example, if the maximum total efficiency is achieved, then the least effort implies the total income of the population should reach its maximum as well. In other words, the two maximums are mutually conditioned through least effort. The functional to be maximized should be the sum $(\eta + c\bar{X})$. We write now

$$\delta(\eta + c\bar{X}) = 0 \tag{15}$$

where $c$ is a constant multiplier characterizing the balance between the two maximums.

## 6) Deriving Pareto law

Eqs.(14) and (15) mean $\frac{\partial}{\partial p_i}\left(\frac{p_i^{1-a}-1}{a} + cp_i x_i\right) = 0$ for all $i$. After the normalization, the result is

$$p_i = C x_i^{-\frac{1}{a}} \tag{16}$$

where the normalization constant $C = 1/\sum x_i^{-\frac{1}{a}}$. Remember that $p_i$ is the probability of finding an agent at the value $x_i$ of $X$.

The continuous version of the discrete distribution Eq.(16) is the probability density function $\rho(x) = Cx^{-\frac{1}{a}}$ with $dp(x) = \rho(x)dx = Cx^{-\frac{1}{a}}dx$ the probability of finding an agent in the interval from $x$ to $x + dx$. The Pareto law follows from the integral of $dp$ from $x$ to the maximum value of $X$ or infinity for simplicity: $p(X > x) = \int_x^\infty Cx^{-\frac{1}{a}}dx = \frac{C}{\frac{1}{a}-1}x^{-(\frac{1}{a}-1)}$. Since $p(X > x_{min}) = 1$ with $x_{min}$ the minimum value of $X$, one gets

$$p(X > x) = \left(\frac{x_{min}}{x}\right)^{\frac{1}{a}-1} \tag{17}$$

which is the Pareto distribution Eq.(2) with $\beta = \frac{1}{a} - 1$. $p(X > x)$ is a decreasing distribution, it follows that $0 < a < 1$ and $0 < \beta < \infty$.

## 7) Deriving Zipf's law

Now let us put the values of $X$ into $W$ bins. These bins are ranked in a decreasing order in the magnitude of $x$. Let $x_r$ be the benchmark value of the bin of rank $r$, we have $x_1 > x_2 ... > x_r > ... > x_W$. The Zipf's law Eq.(1) describes the relationship between $x_r$ and $r$. As mentioned above, the Zipf's law and the Pareto law are regarded as two sides of the same thing and connected to each other by the fact that the cumulate probability $p(X > x)$ is proportional to rank $r$ [19]. But there is also a point of view treating these power laws as two separate laws independent from each other [20]. Here we will show that these two laws are both consequence of MAXEFF and related to each other through another parameter.

By definition, the population (number of agents) having more income than $x_r$ increases with increasing $r$. In other words, $p(X > x_r)$ increases when $r$ increases until its maximum $W$ at which $X$ reaches its minimum value $x_W$ and $p(X > x_W) = 1$. But saying that the probability $p(X > x)$ is always proportional to the rank $r$ seems to be just an observation from some empirical results. It would be hard to say it is a general rule. In what follows, I will use MAXEFF to derive a general relationship between $p(X > x_r)$ and $r$ leading to Zipf's law from Pareto law Eq.(17).

Let us now use the average of the rank in the MAXEFF. Notice that if an agent increases its income $X$, its rank value decreases. Hence whenever $\bar{x} = \sum p_i x_i$ has a maximum, the average rank $\bar{r} = \sum p_r r$ should reach its minimum where $p_r = C x_r^{-\frac{1}{a}}$ is the probability to find an agent of rank $r$ or of the value $x_r$, given by Eq.(16). The calculus of variation applies with $\delta(\eta - c'\bar{r}) = 0$ which should be maximum because $(\eta - c'\bar{r})$ is a difference between the maximum $\eta = \frac{\sum_r p_r^{1-b} - 1}{b}$ ($b$ is not necessarily equal to $a$) and the minimum $\bar{r}$. This leads to $\frac{\partial}{\partial p_r}\left(\frac{p_r^{1-b}-1}{b} - c'p_r r\right) = 0$ and $p_r = C' r^{-\frac{1}{b}}$. By definition of rank distribution, $p_r$ must be increasing function of $r$, hence $b$ must be negative. For simplicity, let $\gamma = -\frac{1}{b} > 0$, we have

$$p_r = C' r^\gamma \qquad (18)$$

Substituting this equation into $p_r = C x_r^{-\frac{1}{a}}$ gives $C' r^\gamma = C x_r^{-\frac{1}{a}}$ and $x_r = \left(\frac{C'}{C}\right)^{-a} r^{-a\gamma}$. Notice that $x_1 = \left(\frac{C'}{C}\right)^{-a}$, Zipf's law reads

$$x_r = \frac{x_1}{r^\alpha} \qquad (19)$$

with $\alpha = a\gamma$. An example of this relationship comes from the Zipf-Pareto distributions of American city populations [6]. The Pareto distribution Eq.(17) shows $\beta = 1.366$, leading to $a = 0.423$. While the Zipf's distribution shows $\alpha = 0.823$, meaning that $\gamma = 1.95$ or $p_r = C'r^{1.95}$ for the system of city population. Their result shows that Zipf and Pareto laws are two different power laws but related to each other by a third power law $p_r \propto r^\gamma$ of the probability to find an agent of rank $r$, as a consequence of MAXEFF. Note that the relationship $\alpha = \frac{1}{\beta+1}\gamma$ between $\alpha$ and $\beta$ is not the same as $\alpha = \frac{1}{\beta}$ obtained with the hypothesis of $p(X > x_r) \propto r$ [6][19].

An extension of the above derivation of Zipf law can be easily made by adding the normalization $\sum p_r = 1$ into the variational calculus as a constraint with one more multiplier $c''$, leading to $\frac{\partial}{\partial p_r}\left(\frac{p_r^{1-b}-1}{b} - c'p_r r - c''p_r\right) = 0$ and $p_r = C'(r+r_0)^\gamma$ through similar mathematical trick, where $r_0 = c''/c'$ is just a new parameter introduced by the normalization constraint. Considering again $p_r = Cx_r^{-\frac{1}{a}}$, we get $C'(r+r_0)^\gamma = Cx_r^{-\frac{1}{a}}$ leading to Zipf-Mandelbrot law

$$x_r = \frac{B}{(r+r_0)^\alpha} \qquad (20)$$

where $\alpha = a\gamma$ and $B = x_1(1+r_0)^\alpha$ a constant for a given rank distribution. Here the problem of the mutual dependence of the two parameters ($\alpha$ and $r_0$) [22] is absent. Eq.(**20**) fits better the data having upper flattening at small rank [18]. Naturally, Zipf law recovers in the special case of $r_0 = 0$.

## 8) Zipf-Pareto efficiency as a measure of performance

The efficiency given by Eq.(14) (from now on we refer to it as Zipf-Pareto or ZP efficiency) provides a possible measure of performance of the ensemble of agents as a whole all making effort for some fulfillment. Using the probability density function $\rho(x) = \frac{1}{Z}x^{-\frac{1}{a}}$, the efficiency reads:

$$\eta = \frac{\int_{x_{min}}^{\infty} \rho^{1-a}\,dx - 1}{a} \qquad (21)$$

The partition function $Z = \int_{x_m}^{\infty} x^{-\frac{1}{a}}dx = \frac{1}{\beta}x_{min}^{-\beta}$. Choosing $x_{min} = 1$, we get $Z = \frac{1}{\beta} = \frac{a}{1-a}$. Since $0 < a < 1$ and $0 < \beta < \infty$, the partition function is increasing function of $a$ in the interval $0 \leq Z < \infty$. The average of $x$ is given by $\bar{x} = \int_{x_{min}}^{\infty} x\rho \, dx = \frac{\beta}{\beta-1} = \frac{1-a}{1-2a}$ which increases with increasing $a$ up to infinity for $a = 0.5$ and becomes negative for $0.5 < a \leq 1$. If we impose the condition of $\bar{x} \geq 0$, then $0 < a \leq 0.5$ and $1 < \beta < \infty$.

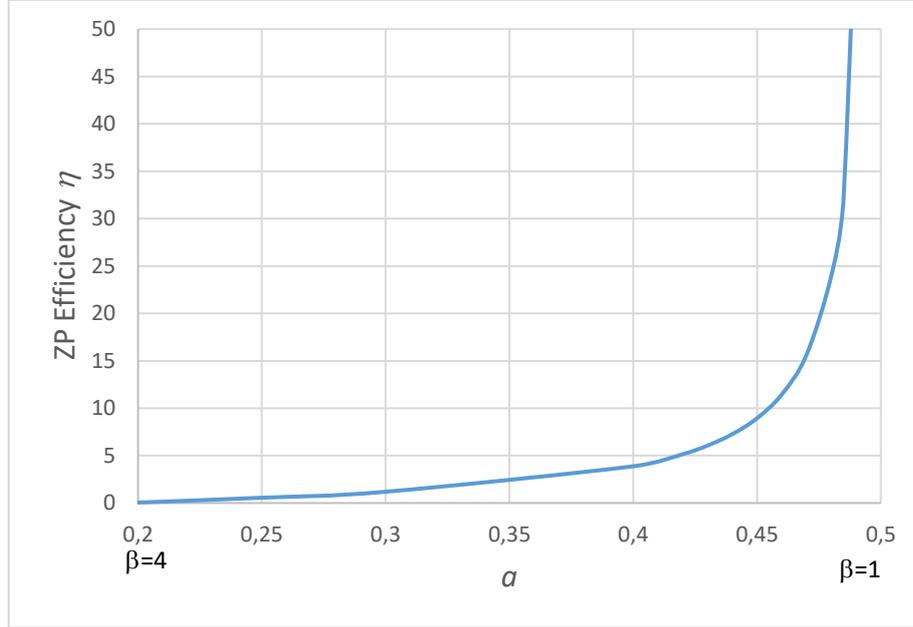

**Figure 2**: Evolution of the ZP efficiency $\eta$ of $\eta = \frac{1}{a}\left[\left(\frac{a}{1-a}\right)^a \frac{1-a}{1-2a} - 1\right]$ as a function of the parameter $a$ in the interval $0.2 < a < 0.5$ or $4 < \beta < 1$. $E$ diverges for $a=0.5$ or $\beta = 1$.

Finally, the integral in Eq.(21) gives $\eta = \frac{Z^a \bar{x}-1}{a} = \frac{1}{a}\left[\left(\frac{a}{1-a}\right)^a \frac{1-a}{1-2a} - 1\right]$ which increases from zero to infinity with increasing $a$ in the interval $0 \leq a < 0.5$. . It is worth noticing that the ZP efficiency increases when $\beta = \frac{1}{a} - 1$ decreases from infinity ($a$=0) to 1 ($a$=0.5). This evolution of the ZP efficiency is plotted in **Figure 2** where only positive efficiency is shown for $\sim 0.2 \leq a$.

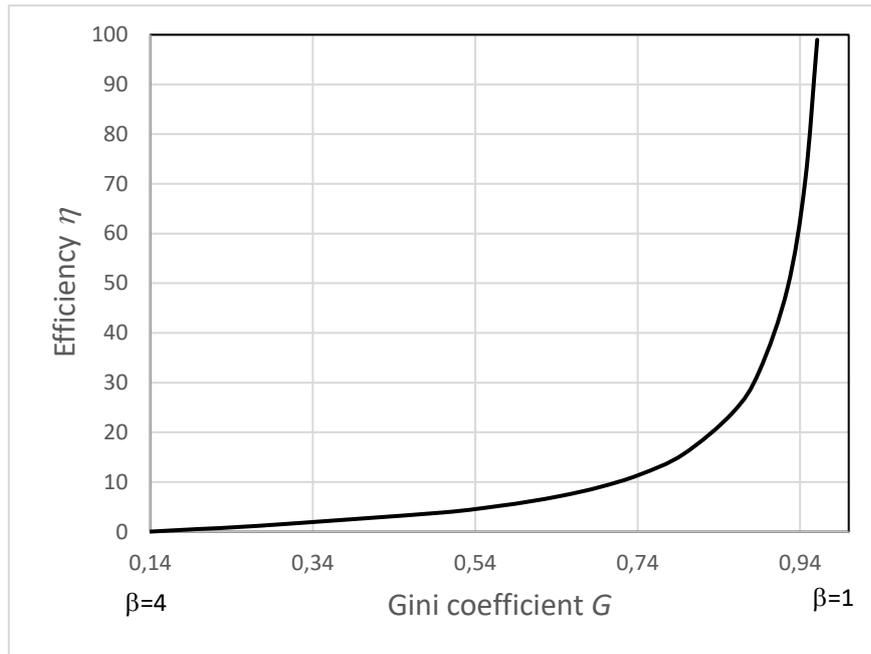

**Figure 3**: Relationship between the ZP efficiency $\eta$ and the Gini coefficient $G$ in the interval of $4 < \beta < 1$. $E$ diverges for $G=1$ or $\beta = 1$.

In order to help the understanding the ZP efficiency, let's compare it to another known properties of Zipf-Pareto distributions, the Gini coefficient $G$, an indicator of inequality for given population in many fields such as economy, sociology, biology, education, health science, ecology and so on[3]. For Pareto distribution, $G = \frac{1}{2\beta - 1}$, meaning that $G = 0$ for $\beta = \infty$ ($a$=0), and $G = 1$ for $\beta = 1$ ($a$=0.5). $G = 0$ implies absolute equality (all agents have the same income for example), and $G = 1$ indicates absolute inequality (one person has all the income and all others have zero income). Clearly, the ZP efficiency increases with increasing $G$ and is very large when $G$ approaches its maximum (see **Figure 3**). This mathematical behavior of the ZP efficiency coincides with the classical point of view in economics [11]. Of course, in a real system, growth may deviate from this ideal behavior and undergoes the impacts of different factors. It is obviously unconvincing to say that an economic system is most efficient when only one hyper-trillionaire has all the wealth and the rest of the population has nothing, or that an educational system has maximal efficiency with only one person very educated and the rest of the population without any education ($G$=1). The relationship between wealth inequality and economic growth is an important issue and widely debated until nowadays[4]. In order to get more insight into the ZP efficiency, it is necessary to make further investigation of its relationship with other properties such as inequality, 80-20 rule, 20/20 ratio, performance, opportunity etc.) by either empirical study or numerical modeling of real systems [12].

---

[3] https://en.wikipedia.org/wiki/Gini_coefficient
[4] https://en.wikipedia.org/wiki/Economic_growth

## 9) Conclusion and remarks

I have implemented the idea of least effort via a calculus of variation MAXEFF in order to derive Zipf-Pareto laws for living systems. I introduce a functional of the efficiency from the consideration of a nonadditive relationship of efficiency of thermal engines in thermodynamics. The Zipf's and Pareto's laws come out naturally from this maximum calculus. This efficiency functional also provides a possible measure of the performance of real systems.

One of the underlying meaning of this approach is that Zipf-Pareto laws are ubiquitous for all systems composed of a large number of living agents, no matter what are their nature and behavior, including human beings, animals or objects recipient of effort of living agents and representing their fulfillment. Whenever a large number of living agents try to achieve something, Zipf-Pareto laws take place in relation with the the output or fulfillment. Regarding objects recipient of effort or fulfillment, one can think about words, webpages, cities, firms, books, phone numbers and so on.

Considering the universal character of the rule of least effort for living systems and that living agents can be likened to thermal engines in term of their efficiency, I hope this approach is meaningful to most, if not all, living systems and systems driven by a large number of living agents statistically obeying least effort and show power law behavior. As example of these systems, we can cite, among others, economic systems, linguistic systems, informational and social networks and so on. These are systems driven by the effort of human beings to achieve something. In economic system regarding income for example, income is the output $X$ people are looking for with effort. People must try in general to achieve more income by trying to do less effort (time, investment, physical effort etc.), leading to Pareto law in the distribution of income [8]. In linguistic systems, similar thing happens since people try not to be talkative (statistically, of course) and to express as much information and idea as possible by using as less words as possible. Here the output should be the quantity of information or ideas expressed, and the effort can be the time spent for writing or speaking or the total number of words as well as the number of different words used. The quantity of information and idea expressed is, of course, not easy to quantify and measure. However, this behavior of trying to express more by using less time and words tends to increase the frequency of individual word in a given text, as if the frequency of words was the thing (output) people are looking for with least effort, leading to Zipf law of frequency-rank distribution. Regarding the development of cities in an economic and political society, people and companies statistically tend to move to large cities for many reasons (financial, communication, education, comfot, entertainment etc.), as if the size of city was something they are trying to obtain with as less effort (time, money, physical effort) as possible, leading to Zipf-Pareto laws of size or distribution. The last example we think worth mentioning is the informational networks where the Zipf-Pareto laws is ubiquitous under the reign of preferential attachment [23]. This connection rule is a typical behavior of least effort or of maximum efficiency in the quest for information or other outputs. Its derivation from least effort will be presented in a coming paper.

We would like to remark the Zipf-Pareto laws are often relevant to Pareto's 80/20 rule[5], or same rule with different proportion, 90/10 for instance). This rule is very helpful for having a glimpse of the ubiquity of Zipf-Pareto type distributions and the underlying least effort and maximum efficiency you can hardly ignore even in daily life. For example, if you have a large number of phone numbers in your calling list, you can easily check that 80% (even more) of your call are made to only 20% (even less) of the phone numbers of the list. Same behavior with your mailing list. If you have a large number of friends, you can easily check a small percentage (perhaps 10%) of your friends occupy most (perhaps 90%) of your time spent with friends. Similar phenomena happen with the books you keep in your library, shoes and cloths if you have many. All this in order to say that least effort should be meaningful for most if not all living systems and systems driven by the effort of large number of living agents who are trying to get more something by doing less.

I used the nonadditivity Eq.(7) or (8) of the efficiency of the heat engines. The reader can notice in the Appendix I that, for heating or cooling engine, similar additivity occurs for the inverse of the efficiency $1/\eta$. This means that $1/\eta$, instead of $\eta$, has a functional similar to Eq.(14). This modification of nonadditivity does not affect the calculus of variation of least effort and maximum efficiency, because if $\eta$ has a maximum, $1/\eta$ should have a minimum, and the calculus of variation $\delta\left(\frac{1}{\eta} + c\bar{X}\right) = 0$ will generate the same probability distribution if $1/\eta$ has the same functional as $\eta$. Therefore, the result of the MAXEFF is independent of the type of the engines in the ensemble.

It is worth noticing that the result of MAXEFF in the present work is a single power law, or Zipf-Mandelbrot law, a near single power law. This is the objective of this work, to find a generic and system independent mechanism, in accordance with the principle of least effort, capable of accounting for the tendency of decreasing rank distribution, a common character to almost all studied data of a large number of very different systems. Once that's done, it would be interesting to consider other mechanisms, which may be system specific, to account for the deviation of single power law. As a matter of fact, most systems claimed to have Zipf-Pareto distributions only show near-Zipf's laws with different (curved down or up) tails or upper/lower cutoff [13][24][25][26][27]. To my opinion, in view of the large number of different deviations observed, it is plausible to consider them system specific and to account for them with specific models, . Within this work, what comes to my mind for the moment is about a double power law, a simple combination of two single power laws, capable of describing very well certain observed data [26]. It is possible to account for this behavior within this framework by considering an ensemble of agents, all differ in behavior as well as in the parameter $a$, in such a way that they can be reorganized into two or more subsystems with different parameter $a$. This composite approach is to be developed in another work by using the technique of incomplete statistics [15]. It is also possible to explain the exponential distributions of certain complex systems [14] since the efficiency functional approaches logarithm form for small $a$. The MAXEFF can generate near-exponential probability distribution in this case.

---

[5] https://en.wikipedia.org/wiki/Pareto_principle

The resemblance of the ZP efficiency to Tsallis entropy[6] [15][16] is noteworthy. This comes from the fact that the ZP efficiency has the same nonadditivity Eq.(7) as Tsallis entropy. However, for Tsallis entropy, this nonadditivity is a consequence of the postulated entropy [16]. But here, the nonadditivity, as suggested by the efficiency of thermal engines, is the starting point which uniquely leads to ZP efficiency. The parameter $a$ has a concrete physical meaning as shown in the Appendix I. The maximization of Tsallis entropy has its origin in the Jaynes principle[7] [17] stipulating the maximization of Boltzmann entropy for thermodynamic systems, while the maximum efficiency arises from the principle of least effort for living systems which is well illustrated by the dictum "achieving more by doing less" that each of us is applying in every detail of our life.

## Acknowledgement

The author thanks Dr Aziz El Kaabouchi, Dr Congjie Ou, Dr Ru Wang, and Dr François-Xavier Machu for valuable discussion and help in the calculation.---

[6] https://en.wikipedia.org/wiki/Tsallis_entropy
[7] https://en.wikipedia.org/wiki/Principle_of_maximum_entropy

## Appendix I

Analysis of the nonadditivity of the efficiency of thermal engines.

**A) Nonadditivity of the efficiency of heat engine doing work**

The definition of efficiency of a thermal engine in thermodynamics differs for one type of engines to another. For example, suppose an engine absorbs an energy $Q_1$, does a useful positive work $W$, and rejects an energy $Q_2$. In the ideal case without energy loss where all heat cost $Q_1 - Q_2$ is converted into work $W$, we have $W = Q_1 - Q_2$. the efficiency of this engine is defined by

$$\eta = \frac{W}{Q_1} = 1 - \frac{Q_2}{Q_1}$$

**A1)** If two engines are connected in such a way that the first engine absorbs an energy $Q_1$, does a work $W_1$, and rejects an energy $Q_2$, and the second engine absorbs the energy $Q_2$, does a work $W_2$, and rejects an energy $Q_3$, one has $\eta_1 = 1 - \frac{Q_2}{Q_1}$, $\eta_2 = 1 - \frac{Q_3}{Q_2}$, and the efficiency $\eta$ of the ensemble of two engines is given by

$$\eta = \frac{W_1 + W_2}{Q_1} = 1 - \frac{Q_3}{Q_1} = 1 - \frac{Q_2}{Q_1}\frac{Q_3}{Q_2} = 1 - (1-\eta_1)(1-\eta_2) = \eta_1 + \eta_2 - \eta_1\eta_2$$

**A2)** Now if the second engine only absorbs a part of the energy $Q_2$, say, $bQ_2$ (b<1), does a work $W_2$, and rejects an energy $Q_3$, one has $\eta_1 = 1 - \frac{Q_2}{Q_1}$, $\eta_2 = \frac{W_2}{bQ_2} = \frac{bQ_2 - Q_3}{bQ_2} = 1 - \frac{Q_3}{bQ_2}$, and the overall efficiency $\eta$ of the ensemble of two engines:

$$\eta = \frac{W_1 + W_2}{Q_1} = \frac{Q_1 - Q_2 + bQ_2 - Q_3}{Q_1} = \frac{(b-1)Q_2 + Q_1 - Q_3}{Q_1} = \frac{(b-1)Q_2}{Q_1} + 1 - \frac{Q_3}{Q_1}$$

$$= (b-1)(1-\eta_1) + 1 - b(1-\eta_1)(1-\eta_2)$$

$$= \eta_1 + b\eta_2 - b\eta_1\eta_2$$

**A3)** Now suppose not all heat $Q_1 - Q_2$ is converted into work $W$ due to some energy loss (friction, thermal radiation, vibration etc.), we can write

$W = \frac{1}{a}(Q_1 - Q_2)$ and $E_1 = \frac{1}{a}\left(1 - \frac{Q_2}{Q_1}\right)$, $E_2 = \frac{1}{a}\left(1 - \frac{Q_3}{Q_2}\right)$,

where $a>1$ characterizes the loss of energy of the engines,

$$\eta = \frac{W_1 + W_2}{Q_1} = \frac{1}{a}\left(1 - \frac{Q_3}{Q_1}\right) = \frac{1}{a}\left(1 - \frac{Q_2}{Q_1}\frac{Q_3}{Q_2}\right)$$

$$= \frac{1}{a}[1 - (1 - a\eta_1)(1 - a\eta_2)]$$

$$= \eta_1 + \eta_2 - a\eta_1\eta_2$$

or
$$(1 - a\eta) = (1 - a\eta_1)(1 - a\eta_2)$$

**A4)** If the two engines have different loss coefficients, say, $a_1$ and $a_2$, we have $\eta_1 = \frac{1}{a_1}\left(1 - \frac{Q_2}{Q_1}\right)$, $\eta_2 = \frac{1}{a_2}\left(1 - \frac{Q_3}{Q_2}\right)$, then

$$\eta = \frac{W_1 + W_2}{Q_1} = \frac{1}{a}\left(1 - \frac{Q_3}{Q_1}\right) = \frac{1}{a}\left(1 - \frac{Q_2 Q_3}{Q_1 Q_2}\right)$$
$$= \frac{1}{a}[1 - (1 - a_1\eta_1)(1 - a_2\eta_2)]$$

One gets
$$(1 - a\eta) = (1 - a_1\eta_1)(1 - a_2\eta_2)$$

**B)** Nonadditivity of the efficiency of heat pump

The definition of efficiency of a heat pump (heating engine) in thermodynamics is the following. Suppose heat pump absorbs a heat $Q_1$, consumes a work $W$, and provides a heat $Q_2$ for heating. We have $W = Q_2 - Q_1$ if all work is converted into heat. Its efficiency is defined by

$$\eta = \frac{Q_2}{W} = \frac{1}{1 - \frac{Q_1}{Q_2}}$$

$$\frac{1}{\eta} = \frac{W}{Q_2} = 1 - \frac{Q_1}{Q_2}$$

**B1)** If two pumps are connected in series in such a way that the first pump absorbs an energy $Q_1$, uses a work $W_1$, and supplies $Q_2$, and the second engine absorbs $Q_2$, consumes a work $W_2$, and supplies $Q_3$, one has $\frac{1}{\eta_1} = 1 - \frac{Q_1}{Q_2}$, $\frac{1}{E_2} = 1 - \frac{Q_2}{Q_3}$, and the overall efficiency $E$ of the ensemble of two engines:

$$\frac{1}{\eta} = \frac{W_1 + W_2}{Q_3} = 1 - \frac{Q_1}{Q_3} = 1 - \frac{Q_1 Q_2}{Q_2 Q_3}$$
$$= 1 - \left(1 - \frac{1}{\eta_1}\right)\left(1 - \frac{1}{\eta_2}\right)$$
$$= \frac{1}{\eta_1} + \frac{1}{\eta_2} - \frac{1}{\eta_1}\frac{1}{\eta_2}$$

or

$$\left(1-\frac{1}{\eta}\right)=\left(1-\frac{1}{\eta_1}\right)\left(1-\frac{1}{\eta_2}\right)$$

**B2)** If not all the work $W$ is converted into heat $Q_1 - Q_2$ due to some loss, let $Wa = (Q_1 - Q_2)$

$$\eta_1 = \frac{Q_2}{W_1} = a\frac{Q_2}{Q_2 - Q_1} = \frac{a}{\left(1-\frac{Q_1}{Q_2}\right)}$$

$$\frac{1}{\eta_1} = \frac{1}{a}\left(1-\frac{Q_1}{Q_2}\right)$$

and

$$\frac{1}{\eta_2} = \frac{1}{a}\left(1-\frac{Q_2}{Q_3}\right)$$

where a<1 characterizes the loss of heat energy of the engines.

$$\frac{1}{\eta} = \frac{W_1 + W_2}{Q_3} = \frac{1}{a}\left(1-\frac{Q_1}{Q_3}\right)$$

$$= \frac{1}{a}\left(1-\frac{Q_2 Q_3}{Q_1 Q_2}\right) = \frac{1}{a}\left[1-\left(1-a\frac{1}{\eta_1}\right)\left(1-a\frac{1}{\eta_2}\right)\right]$$

$$= \frac{1}{\eta_1} + \frac{1}{\eta_2} - a\frac{1}{\eta_1}\frac{1}{\eta_2}$$

or

$$\left(1-a\frac{1}{\eta}\right)=\left(1-a\frac{1}{\eta_1}\right)\left(1-a\frac{1}{\eta_2}\right)$$

**B3)** If now the two pumps have different loss coefficients, say, $a_1$ and $a_2$, we have

$$\frac{1}{\eta_1} = \frac{1}{a_1}\left(1-\frac{Q_1}{Q_2}\right)$$

$$\frac{1}{\eta_2} = \frac{1}{a_2}\left(1-\frac{Q_2}{Q_3}\right)$$

then

$$\left(1-a\frac{1}{\eta}\right)=\left(1-a_1\frac{1}{\eta_1}\right)\left(1-a_2\frac{1}{\eta_2}\right)$$

**C) Nonadditivity of the efficiency of refrigerator**

The definition of efficiency of a refrigerator (cooling engine) is the following. Suppose a refrigerator absorbs a heat $Q_1$, consumes a work $W$, and rejects a heat $Q_2$ for cooling. We have $W = Q_2 - Q_1$ if all work is converted into heat. Its efficiency is defined by

$$\eta = \frac{Q_1}{W} = \frac{1}{\frac{Q_2}{Q_1} - 1}$$

$$\frac{1}{\eta} = \frac{W}{Q_1} = \frac{Q_2}{Q_1} - 1$$

**C1)** If two refrigerators are connected in such a way that the first one absorbs an energy $Q_1$, uses a work $W_1$, and rejects $Q_2$, and the second one absorbs $Q_2$, consumes a work $W_2$, and rejects $Q_3$, one has $\frac{1}{E_1} = \frac{Q_2}{Q_1} - 1, \frac{1}{E_2} = \frac{Q_3}{Q_2} - 1$, and the overall efficiency $E$ of the ensemble of two engines reads:

$$\frac{1}{\eta} = \frac{W_1 + W_2}{Q_1} = \frac{Q_3}{Q_1} - 1 = \frac{Q_3}{Q_2}\frac{Q_2}{Q_1} - 1$$

$$= \left(\frac{1}{\eta_1} + 1\right)\left(\frac{1}{\eta_2} + 1\right) - 1$$

$$= \frac{1}{\eta_1} + \frac{1}{\eta_2} + \frac{1}{\eta_1}\frac{1}{\eta_2}$$

or

$$\left(\frac{1}{\eta} + 1\right) = \left(\frac{1}{\eta_1} + 1\right)\left(\frac{1}{\eta_2} + 1\right)$$

**C2)** In case of loss with a coefficient $a$, we have $\eta_1 = \frac{1}{a}\left(\frac{Q_2}{Q_1} - 1\right), E_2 = \frac{1}{a}\left(\frac{Q_3}{Q_2} - 1\right)$,

$$\frac{1}{\eta} = \frac{1}{\eta_1} + \frac{1}{\eta_2} + a\frac{1}{\eta_1}\frac{1}{\eta_2}$$

$$\left(\frac{a}{\eta} + 1\right) = \left(\frac{a}{\eta_1} + 1\right)\left(\frac{a}{\eta_2} + 1\right)$$

**C3)** In case of loss with two different coefficients $a_1$ and $a_2$, we have $\eta_1 = \frac{1}{a_1}\left(\frac{Q_2}{Q_1} - 1\right), \eta_2 = \frac{1}{a_2}\left(\frac{Q_3}{Q_2} - 1\right)$, one gets

$$\left(\frac{a}{\eta}+1\right) = \left(\frac{a_1}{\eta_1}+1\right)\left(\frac{a_2}{\eta_2}+1\right)$$

## Appendix II

The unicity of Eq.(12), $(1 + a\eta_i) = p_i^b$, given Eqs.(10) and (11), i.e., $p_i = p_k(A)p_j(B)$ and $(1 + a\eta_i) = [1 + a\eta_k(A)][1 + a\eta_j(B)]$, can be proved as follows.

**Theorem.** Let $f$ be nonnegative and continuous function on $]0,+\infty[$, and satisfy the condition: For all $x, y \in ]0,+\infty[$, $f(xy) = f(x)f(y)$, then, there exists $b > 0$ such that: for all $x \in ]0,+\infty[$, $f(x) = x^b$.

*Proof.*

Putting $g(x) = \ln(f(e^x))$, $g$ is continuous on $\mathbf{R}$, and for all $x, y \in \mathbf{R}$, we have

$$\begin{aligned} g(x+y) &= \ln(f(e^{x+y})) \\ &= \ln(f(e^x e^y)) \\ &= \ln(f(e^x)f(e^y)) \\ &= \ln(f(e^x)) + \ln(f(e^y)) \\ &= g(x) + g(y) \end{aligned}$$

Consequently, for all $n \in \mathbf{N}^*$, we have $g\left(\dfrac{1}{n}\right) = \dfrac{1}{n}g(1)$, because

$$g(n) = g(1+1+\cdots+1) = ng(1) \text{ and } g(1) = g\left(\dfrac{1}{n}+\dfrac{1}{n}+\cdots+\dfrac{1}{n}\right) = ng\left(\dfrac{1}{n}\right).$$

For all $p, q \in \mathbf{N}^*$, we have $g\left(\dfrac{p}{q}\right) = \dfrac{p}{q}g(1)$, because

$$g\left(\dfrac{p}{q}\right) = g\left(\dfrac{1}{q}+\dfrac{1}{q}+\cdots+\dfrac{1}{q}\right) = pg\left(\dfrac{1}{q}\right) = p\dfrac{1}{q}g(1).$$

Now, if $x \in ]0,+\infty[$, there exists a sequence $(p_n, q_n)$ in $\mathbf{N}^* \times \mathbf{N}^*$, such that $x = \lim\limits_{n \mapsto +\infty} \dfrac{p_n}{q_n}$.

Because $g$ is continuous on $]0,+\infty[$, we have

$$g(x) = \lim_{n \mapsto +\infty} g\left(\dfrac{p_n}{q_n}\right) = \lim_{n \mapsto +\infty} \dfrac{p_n}{q_n} g(1) = xg(1).$$

Conclusion, for $x \in ]0,+\infty[$, $g(x) = xg(1)$. For $x \in ]0,+\infty[$, we put $X = \ln x$ and $a = g(1) = f(e^1)$, we have

$$f(x) = f(e^X) = e^{g(X)} = e^{Xg(1)} = x^{g(1)} = x^b.$$

Let $f(x) = 1 + a\eta_i(x)$, Eq.(12) is proved.